\documentclass[notoc]{JHEP3}

\usepackage{epsfig}

\newcommand{\beq}{\begin{equation}}
\newcommand{\eeq}{\end{equation}}
\newcommand{\bea}{\begin{eqnarray}}
\newcommand{\eea}{\end{eqnarray}}
\newcommand{\nn}{\nonumber}

\def\eqn#1{Eq.~(\ref{#1})}
\def\eqns#1#2{Eqs.~(\ref{#1}) and~(\ref{#2})}
\def\eqnss#1#2{Eqs.~(\ref{#1})-(\ref{#2})}
\def\fig#1{Fig.~{\ref{#1}}}
\def\sec#1{Section~{\ref{#1}}}
\def\app#1{Appendix~\ref{#1}}

\newcommand\sss{\scriptscriptstyle}
\def\ord{{\cal O}}
\def\Qb{\bar{Q}}
\def\as{\alpha_{\sss S}}
\def\gs{g_{\sss S}}

\newcommand\fverb{\setbox\pippobox=\hbox\bgroup\verb}
\newcommand\fverbdo{\egroup\medskip\noindent%
                        \fbox{\unhbox\pippobox}\ }
\newcommand\fverbit{\egroup\item[\fbox{\unhbox\pippobox}]}
\newbox\pippobox
\title{Heavy-quark production at large rapidities at hadron colliders}

\author{Jeppe R.~Andersen\\
Cavendish Laboratory, University of Cambridge,\\ 
Madingley Road, CB3 0HE, Cambridge, UK, and\\
DAMTP,  Centre for Mathematical Sciences, \\
Wilberforce Road, CB3 0WA, Cambridge, UK\\
E-mail: \email{andersen@hep.phy.cam.ac.uk}}

\author{Vittorio Del Duca\\
Istituto Nazionale di Fisica Nucleare, Sez. di Torino\\
via P. Giuria, 1 - 10125 Torino, Italy\\
E-mail: \email{delduca@to.infn.it}}

\author{Stefano Frixione\\
Istituto Nazionale di Fisica Nucleare, Sez. di Genova\\
Via Dodecaneso 33, 16124 Genova, Italy\\
E-mail: \email{Stefano.Frixione@cern.ch}}

\author{Fabio Maltoni\\
Centro Studi e Ricerche ``Enrico Fermi'', \\
via Panisperna, 89/A - 00184 Rome, Italy, and\\
Dipartimento di Fisica, Terza Universit\`a di Roma, \\
via della Vasca Navale, 84 - 00146 Rome, Italy\\
E-mail: \email{maltoni@fis.uniroma3.it}}

\author{W. James Stirling\\
Institute for Particle Physics Phenomenology\\
University of Durham - Durham, DH1 3LE, U.K.\\
E-mail: \email{W.J.Stirling@durham.ac.uk}}

\abstract{We investigate heavy-quark production as a function of
the rapidity interval between two heavy quarks in hadronic collisions.
We compare the results relevant to bottom production at the Tevatron 
and at LHC, obtained using exact leading-order
and NLO pQCD production, as well as the contribution of the $4b$
channel with and without the addition of BFKL gluon radiation.}

\preprint{
{Cavendish-HEP-04-26,~DAMTP-2004-83},
{~DFTT 20/2004},
{~GEF-TH-10/2004},
{~RM3-TH/04-19}\\
{~DCPT/04/98,~IPPP/04/49}
}

\begin{document}

\section{Introduction}
\label{sec:intr}

One of the most important processes at high-energy hadron-hadron colliders is
the production of heavy quarks. Bottom and top quark production, for example,
provide not only many tests of perturbative QCD, but also some of the most
important backgrounds to new physics processes. Not surprisingly, therefore,
such heavy-quark production has been extensively studied in the literature
(see e.g. Ref.~\cite{Frixione:1997ma} for a review) and the phenomenology 
at the Tevatron and the LHC has been evaluated in great detail.

In the kinematic region in which the transverse momentum of the heavy
quark $Q$ is of the same order as its mass $m_Q$, 
the leading-order contribution to the inclusive heavy-quark production cross
section comes from the partonic subprocesses in which a $Q\bar{Q}$ pair is
produced, $gg, q \bar q \to Q \bar{Q}$. The next-to-leading order (NLO) 
corrections to these processes have been available for quite some 
time now~\cite{Nason:1987xz,Beenakker:1990ma,Nason:1989zy,Mangano:jk}.
They are numerically important, particularly for $b$ quarks, where they 
can result in a $K$ factor as large as two.

At the parton level, these large radiative corrections to the total rates
are easily identified as coming from production near threshold, 
$\hat{s}\sim 4 m_Q^2$ ($\hat{s}$ being the partonic centre-of-mass
energy squared). When folding partonic cross sections with parton
distribution functions (pdfs) to get the observable rates, the threshold
region is especially relevant in those cases in which the total hadronic energy
$\sqrt{S}$ is of the same order as the quark mass, as for example for
top production at the Tevatron, or $b$ production at fixed-target facilities.
Potentially large logarithms appear in the perturbative expansion,
and these need to be resummed to all orders. In practice, however,
this resummation only marginally increases the NLO predictions
(see e.g. Ref.~\cite{Bonciani:1998vc}).

Total partonic rates can also receive large contributions from the
high-energy region $\hat{s}\gg 4 m_Q^2$, complementary to the threshold
region. As discussed in Ref.~\cite{Nason:1987xz}, this is due to those
partonic subprocesses that feature a gluon exchange in the $t$-channel; this
happens for $gg\to Q\bar{Q}g$ and $qg\to Q\bar{Q}q$, and it is peculiar to
the NLO computations of quark pair production, as opposed to Born-level
predictions, in which only fermions are exchanged in the $t$ channel. It must
be stressed that at the hadron level this enhancement is diluted by the
fall-off of the pdfs at large $x$ values~\cite{DelDuca:mn,Stirling:1994zs}.

A gluon exchange in the $t$-channel is also present at ${\cal O}(\as^4)$
in the reaction $gg\to Q\bar{Q}Q\bar{Q}$, which is the Born-level
contribution to this four-quark process. This is interesting, since the $t$-channel
gluon exchange leads to 
properties fairly similar to those relevant to the 
\emph{Mueller-Navelet} dijet cross section~\cite{Mueller:ey},
which is used to study the high-energy limit of QCD in which the 
energy dependence of the lowest-order cross section is enhanced by 
BFKL-type logarithmic corrections~\cite{Kuraev:ge,Kuraev:1977fs,Balitsky:1978ic}. 

The dominance of the gluon exchange in the $t$-channel implies that 
the $4Q$ channel is perturbatively suppressed only by a factor of $\as$ 
with respect to pair production at high energies. 
Although this still prevents us from a
straightforward use of $4Q$ production to detect BFKL
signals, we can, however, observe that in the high-energy regime
the kinematics of the $2Q$ and $4Q$ production channels are rather different.
The former is dominated by those configurations in which the $Q\bar{Q}$ 
pair recoils with large rapidity against a fast light parton. 
On the other hand, the $4Q$ system will predominantly be produced in two
$Q\bar{Q}$ pairs, rapidly moving away from each other; the relative 
rapidity of each pair is small compared to the separation in rapidity 
of the two pairs. Therefore by selecting particular kinematic configurations
it may be possible to relatively enhance the $4Q$ contribution and find signatures
of BFKL. This is the main focus of our study.

In order to define a proper set of observables, we require for each
event to tag (at least) two heavy flavours (in any possible combination:
$Q\bar{Q}$, $QQ$, or $\bar{Q}\bar{Q}$), which we denote by $Q_1$ and $Q_2$,
separated by a large rapidity interval, $\Delta y=|y_{Q_1}-y_{Q_2}|\gg 1$.
In this way, we should cut off the configurations that dominate pair
production in the high-energy regime, without losing too many events
in the $4Q$ channel. We aim at studying whether this is the case or not, specifically
in the regions accessible to the detectors at present and future colliders,
by comparing the predictions for $2Q$ and $4Q$ production processes.
We stress that our set of observables is based on a double $Q$ tagging,
which in fact is already used  to study $Q\bar{Q}$ correlations 
in heavy-quark pair production. In this paper, we shall not correct our results for
tagging efficiency.
% thus, our conclusions should be considered as an
%upper theoretical bound to the search for BFKL signals in heavy-flavour
%production at hadron colliders.

The high-energy limit of $4Q$ production
can be considered as a reformulation of the standard
Mueller-Navelet dijet case. What we are doing here, in effect, is replacing 
each Mueller-Navelet forward jet (with $p_T > P_{T{\rm min}}$) by a 
$Q\bar Q$ pair. In fact, by identifying the rapidity of each pair 
with the rapidity of the tagged quark in the pair, we have
$\hat{s}=4 m_{Q_\perp}^2 \cosh^2 y^*$, where 
$m_{Q_{\perp}}^2 = {p^2_{Q_{\perp}} + m_Q^2}$ is the squared
heavy-quark transverse mass, and $y^* = (\Delta y)/2$. The formula
above, relating the large-$\hat{s}$ to the large-$y^*$ region, is customary
in Mueller-Navelet arguments. The differences between jet and heavy-quark 
production are easy to find: whereas in the dijet case it is 
$P_{T{\rm min}}$ that regulates the infrared singularities at $\hat{t}=0$,
here it is the heavy-quark mass $m_Q$. The analogue of the $P_{T{\rm min}}^2
\hat\sigma_{jj} \to\ $constant behaviour of the leading-order dijet cross
section at large dijet rapidity separation $\Delta y$ is the $m_Q^2
\hat\sigma_{4Q} \to\ $constant behaviour of the $4Q$ heavy-quark cross
section. The effect of the (leading logarithm) BFKL corrections is the same in
both cases: the partonic cross sections increase asymptotically as 
$\exp(\lambda \Delta y)$ where 
\textbf{$\lambda=4\log{2} N_c\as/\pi$} and $\Delta y$ is either the
rapidity separation of the dijets in the Mueller-Navelet case, or the rapidity
separation of the two $Q\bar Q$ systems in the present context.

Another process of potential interest in the high-energy limit is
$Q\bar{Q} + 1$~jet production. In this case the partonic subprocesses
$gg\to Q\bar{Q}g$ and $qg\to Q\bar{Q}q$, which feature a gluon exchange in 
the $t$-channel, are $\ord(\as^3)$ at the Born level. This can also be
considered as a reformulation of the standard Mueller-Navelet analysis,
where only one of the forward jets is replaced by a $Q\bar{Q}$ pair.

This paper is organized as follows. In Section~\ref{sec:hel} we
compute analytically the high-energy limit of the $gg\to Q\Qb Q\Qb$
cross section. In Section~\ref{sec:bfklmc}, we describe how to 
include the resummation of BFKL logarithms, through
Monte Carlo methods. In Section~\ref{sec:bfklobs}, results for 
$2Q$ and $4Q$ channels are compared, at the Tevatron and LHC energies.
We also consider the case of $Q\bar{Q} + 1$~jet production.
Finally, in Section~\ref{sec:conc} we present our conclusions. The
appendices collect some useful formulae.

\section{The high-energy limit}
\label{sec:hel}

In the high-energy limit, the $\Delta y$ distribution for $Q\bar{Q}$
production can be written schematically as
\beq
\frac{d\sigma_{Q\bar Q}}{\Delta y}
\,\sim\, \as^2 \sum_{j=0}^\infty a_{0j} \as^j\
+\  \as^4 \sum_{j=0}^\infty a_{1j} (\as L)^j\ +\  \as^4 \sum_{j=0}^\infty
a_{2j} \as(\as L)^j\ + \cdots,
\label{BFKLxsec}
\eeq
where \mbox{$L=\log(\hat s/\mu_{\sss\rm W}^2) \simeq\Delta y$} 
is a large logarithm, and the quantity $\mu_{\sss\rm W}^2$ is a
mass scale squared, typically of the order of the crossed-channel momentum
transfer and/or of the heavy-quark masses.
The first sum in \eqn{BFKLxsec}
is a fixed-order expansion in $\as$ starting at $\ord(\as^2)$ (the Born processes
$q\bar q, gg \to Q\bar Q$), which
collects together the contributions that do not feature gluon exchange in the
crossed channel between the heavy quarks.
\EPSFIGURE[t]{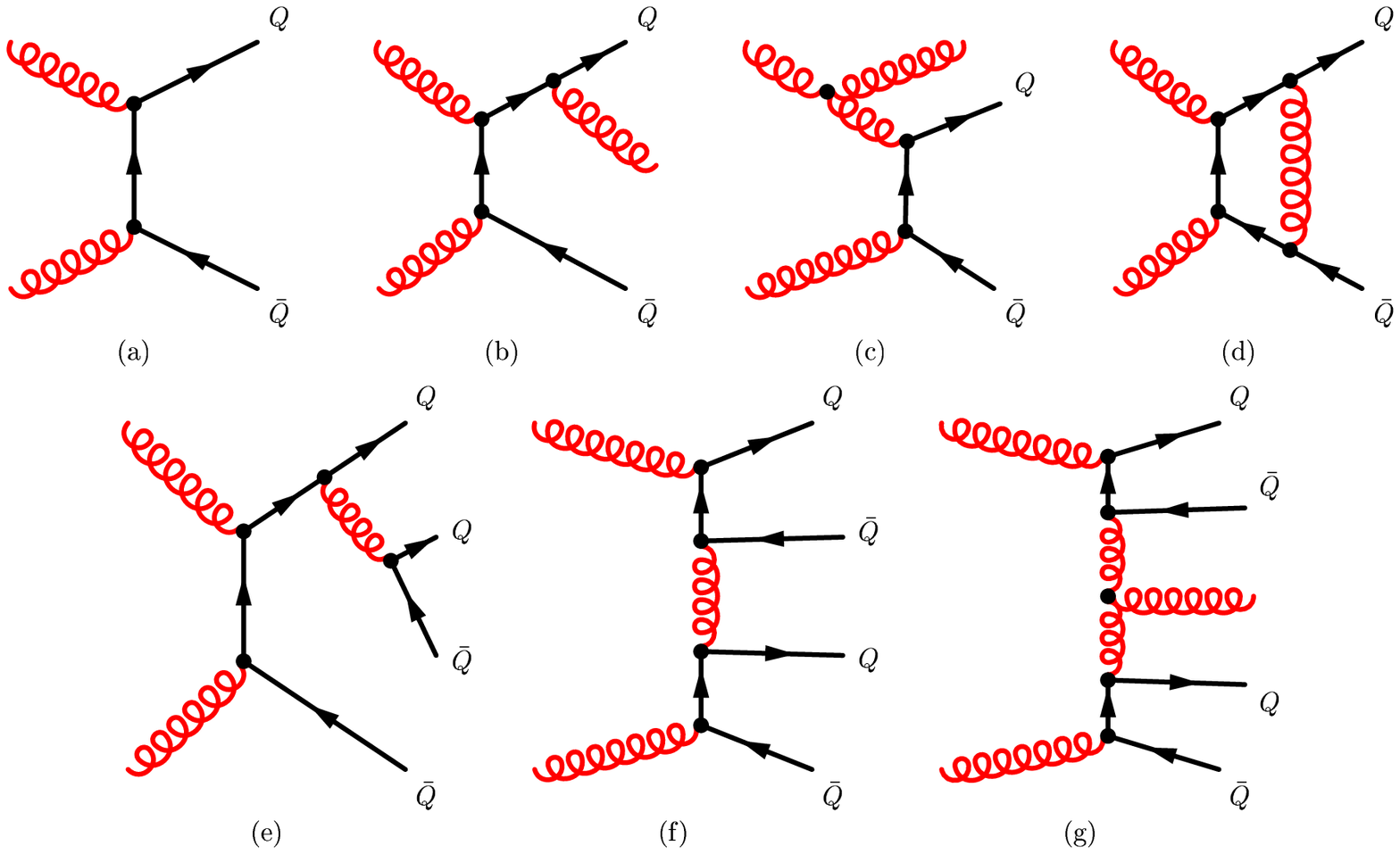,width=15cm}
{Amplitudes for $Q\bar Q$ production in $g g$ fusion.
Figure (a) represents the leading-order term. Figures (b), (c) and
(d) are examples of the diagrams that contribute to the NLO term. 
Figures (e) and (f) represent the 4~$Q$ contribution to the NNLO term; 
figure (e) ((f)) is an example of a diagram with quark (gluon) 
exchange in the $t$ channel. Figure (f) constitutes also the leading term
of a BFKL gluon ladder, and figure (g) represents the first rung of it.
\label{fig:bb}  }
The $a_{00}$ coefficient is the leading-order term, which for $gg$
fusion is depicted in \fig{fig:bb}(a); the $a_{01}$ coefficient is the
NLO term (specimen diagrams are given in \fig{fig:bb}(b-d)). An
example of a $4Q$ contribution to the $a_{02}$ coefficient is given in
\fig{fig:bb}(e).  The $a_{0j}$ coefficients behave like $1/\hat s$, or
equivalently $\exp(-\Delta y)$, modulo logarithmic
corrections.\footnote{The $a_{02}$ coefficient may also contain terms
that behave like $1/(\sqrt{\hat s} \mu_{\sss\rm W})$ and arise from
the interference between diagrams with gluon exchange in the crossed
channel and diagrams with quark exchange in the crossed channel.} In
\eqn{BFKLxsec}, the second and third sums collect the contributions
which feature \emph{only} gluon exchange in the crossed channel
between heavy quarks, the second (third) sum resumming the BFKL
(next-to-)leading logarithmic corrections. \fig{fig:bb}(f) represents
the zeroth-order term, and \fig{fig:bb}(g) contributes to the
first-order term, of the second sum. The $a_{1j}$ and $a_{2j}$
coefficients behave like $1/\mu_{\sss\rm W}^2$, in contrast to the
$1/\hat s$ behaviour of the $a_{0j}$.  The ellipses of \eqn{BFKLxsec}
refer to logarithmic corrections beyond the next-to-leading accuracy.
Thus, it is clear that the second and third sums of \eqn{BFKLxsec}
will eventually dominate over the first sum in the asymptotic energy
region $\hat{s}\to\infty$. In Sections~\ref{sec:hel} and
\ref{sec:bfklmc} we will analyse the second sum of \eqn{BFKLxsec}
 in the region $\hat{s}\gg\mu_{\sss\rm W}^2$, by computing
the $a_{1j}$ coefficients in the high-energy
limit.\footnote{Contributions like the one in \fig{fig:bb}(c), which
feature gluon exchange in the crossed channel but not between heavy
quarks, are not systematically resummed in \eqn{BFKLxsec}, and are
thus implicitly included in the first sum. They contribute, however,
to the leading order for $Q\bar{Q} + 1$~jet production, where a gluon
is exchanged in the $t$-channel between the jet and the $Q\bar{Q}$
pair, and they constitute in that case the Born term of the BFKL
ladder. We will consider $Q\bar{Q} + 1$~jet production in
\sec{sec:bfklobs2q1j}.}

Details of the calculation for the production of four heavy quarks, 
via the sub-processes $g g\to Q\Qb Q \Qb$ and $q \bar{q}\to Q
\Qb Q \Qb$, are presented in \app{sec:4heavy}. 
In the high-energy limit, we require that any two $Q$'s (no distinction
between $Q$ and $\Qb$ is necessary) are produced at large rapidity
separation. Then the production process is dominated by the sub-processes for 
which the tagged $Q$'s are separated by gluon exchange in the crossed channel.
Of the above two sub-processes, only $g g\to Q\Qb Q \Qb$ features gluon 
exchange in the crossed channel. With the kinematics of the
high-energy limit,
\begin{equation}
y_{Q_1}\simeq y_{\Qb_2} \gg y_{Q_3}\simeq y_{\Qb_4} \, ,\qquad
p_{Q_{1\perp}} \simeq p_{\Qb_{2\perp}} \simeq
p_{Q_{3\perp}} \simeq p_{\Qb_{4\perp}}\, ,\label{hekin}
\end{equation}
the amplitude for $g g\to Q \Qb Q \Qb$ factorises as 
\begin{equation}
|{\cal M}_{g_a g_b \to Q_1 \Qb_2 Q_3 \Qb_4 }|^2 =
\frac{4 \hat{s} ^2}{\hat{t} ^2} \left[
                  I^{Q\Qb}(p_a,p_{Q_f},p_{{\Qb}_f};q)
                  I^{Q\Qb}(p_b,p_{Q_b},p_{{\Qb}_b};-q)\right]\, ,
\label{ggQQQQfact}
\end{equation}
where $Q_f (\Qb_f) $ and $Q_b (\Qb_b)$ are the quarks (anti-quarks) produced
forward and backward, respectively. In \eqn{ggQQQQfact},
\begin{equation}
\hat t = q^2 = (p_a-p_{Q_f}-p_{\Qb_f})^2
\label{hattdef}
\end{equation}
is the momentum transfer. The impact factor $I^{Q\Qb}$ is
calculated in \app{sec:ifqq}, starting from the amplitude for
$g\, i\to Q \Qb  i$ with $i=q,g$ and using high-energy factorisation.
The result is given in \eqn{QQimp},
summed (averaged) over final (initial) colours and helicities.
In the kinematics of (\ref{hekin}), the exact parton 
momentum fractions (\ref{mtmfr}) % become
are well approximated by
\begin{equation}
x_a^0 = {m_{Q_{1\perp}} e^{y_{Q_1}} + m_{\Qb_{2\perp}} e^{y_{\Qb_2}}
\over \sqrt{S} }\; , \qquad
x_b^0 = {m_{Q_{3\perp}} e^{-y_{Q_3}} + m_{\Qb_{4\perp}} e^{-y_{\Qb_4}}
\over \sqrt{S} }\, .\label{hemtmfr}
\end{equation}
Using \eqn{ggQQQQfact},
we can write the cross section for heavy-quark production as
\begin{eqnarray}
\lefteqn{
{d\sigma\over \prod_{i=1}^4 d^2{\bf p}_{Q_{i\perp}} dy_{Q_i} } } \nn\\
&=& x_a^0 g_{a/A}(x_a^0,\mu_{{\sss F}a}^2)\, 
x_b^0 g_{b/B}(x_b^0,\mu_{{\sss F}b}^2)\
{I^{Q\Qb}(q_a) I^{Q\Qb}(q_b) \over 2\pi^4 (4\pi)^4
q_{a_\perp}^2 q_{b_\perp}^2 }\,
{\delta^2( {\bf q}_{a_\perp} - {\bf q}_{b_\perp} )\over 2}
\, ,\label{xsec2}
\end{eqnarray}
with momentum transfers $q_a= p_a - p_{Q_1} - p_{\Qb_2}$ and
$q_b = p_{Q_3} + p_{\Qb_4} - p_b$, and where $g_{a/A}(x_a^0,\mu_{{\sss F}a}^2)$
is the pdf for the gluon $g_a$, and analogously for $g_b$.
We use the notation $\mathbf{p}_\perp$ to denote a transverse momentum vector.

However, in \eqn{xsec2} energy and longitudinal momentum are not conserved.
The parton momentum fractions in the high-energy limit, $x_a^0$ and $x_b^0$,
underestimate the exact ones, $x_a$ and $x_b$, \eqn{mtmfr}
and accordingly the values of the pdfs are overestimated. Thus for the numerical
applications of \sec{sec:bfklobs} we will use the 
factorised form (\ref{xsec2}) of the production rate, but with
$x_a^0\to x_a$ and $x_b^0\to x_b$.
This modification is particularly important when BFKL evolution is considered.

The above results must be integrated over the phase space of the final-state
particles in order to get physical results. In the high-energy limit, 
the phase space (\ref{hadphase}) can be factorised
into the phase spaces for the two impact factors,
\begin{eqnarray}
d{\cal P}_4 &=& \left( \prod_{i=1,2} {d^3 p_{Q_i}\over (2\pi)^3 2p^0_{Q_i}}\,
2\pi \, \delta(p_a^+ - p_{Q_1}^+ - p_{\Qb_2}^+) \right)\, \left( \prod_{i=3,4}
{d^3 p_{Q_i}\over (2\pi)^3 2p^0_{Q_i}}\, 2\pi \,
\delta(p_b^- - p_{Q_3}^- - p_{\Qb_4}^-) \right) \nonumber\\ &&\times
(2\pi)^2 \, \delta^2( {\bf p}_{Q_{1\perp}} + {\bf p}_{\Qb_{2\perp}} +
{\bf p}_{Q_{3\perp}} + {\bf p}_{\Qb_{4\perp}})\, ,\label{heps}
\end{eqnarray}
where we have used light-cone coordinates $p^\pm = (p^0\pm p^3)/\sqrt{2}$.
Fixing
\begin{equation}
z_a = {p_{Q_1}^+\over p^+_{Q_1} + p_{\Qb_2}^+}\, ,\qquad
z_b = {p_{Q_3}^-\over p^-_{Q_3} + p_{\Qb_4}^-}\, ,
\end{equation}
the phase space (\ref{heps}) can be rewritten as
\begin{eqnarray}
d{\cal P}_4 &=& {1\over (4\pi)^2} {1\over 2\hat s}
\left( {dz_a\over z_a(1-z_a)} {d^2{\bf p}_{Q_{1\perp}} \over (2\pi)^2} \right)
\left( {dz_b\over z_b(1-z_b)} {d^2{\bf p}_{Q_{3\perp}} \over (2\pi)^2} \right)
 \nn\\ && \times
{d^2{\bf q}_{a_\perp} \over (2\pi)^2} {d^2{\bf q}_{b_\perp} \over (2\pi)^2}
(2\pi)^2 \, \delta^2({\bf q}_{a_\perp} - {\bf q}_{b_\perp})\,, \label{bfklps}
\end{eqnarray}
with centre-of-mass energy
$\hat s = 2p_a^+ p_b^-$. Note that \eqn{bfklps} is written in such a way as to be
immediately generalizible to the emission of a BFKL gluon ladder
between the impact factors.

Using \eqns{ggQQQQfact}{bfklps} in the expression for the cross section given in (\ref{genps}),
we obtain
\begin{equation}
d{\hat \sigma}(p_a p_b\to  p_{Q_1}, p_{\Qb_2} p_{Q_3} p_{\Qb_4}) =
{d^2{\bf q}_{a_\perp} \over (2\pi)^2} {d^2{\bf q}_{b_\perp} \over (2\pi)^2}
{{\cal I}({\bf q}_{a_\perp}) \over q_{a_\perp}^2}
{{\cal I}({\bf q}_{b_\perp}) \over q_{b_\perp}^2}
(2\pi)^2 \, \delta^2({\bf q}_{a_\perp} - {\bf q}_{b_\perp})\,
,\label{factxsec}
\end{equation}
where the integrated impact factor is
\begin{equation}
{\cal I}({\bf q_{\perp}}) = {1\over 4\pi} \int_0^1 {dx\over x(1-x)}
{d^2{\bf p_{\perp}} \over (2\pi)^2} I^{Q\Qb}(x,{\bf p_{\perp}};
{\bf q_{\perp}})\, ,\label{intif}
\end{equation}
with $I^{Q\Qb}$ given in \eqn{QQimp}.
The integral is explicitly performed in \sec{sec:appa1}, where
it is expressed in terms of a function $g$, \eqn{intif2}, of
the dimensionless ratio $\xi= q_\perp^2/m_Q^2$.
Then using \eqnss{intif0}{intif2}, the
total integrated cross section (\ref{factxsec}) becomes
\begin{equation}
\sigma_{gg} = \frac{\as^4}{m_Q^2} \frac{N_c^2-1}{4 \pi}
\int_0^\infty \; \frac{d\xi_a}{\xi_a} \; \frac{d\xi_b}{\xi_b} \;
g(\xi_a) g(\xi_b) \delta(\xi_a-\xi_b)\, .\label{ggsigma}
\end{equation}
Note that even though according to \eqn{intif1} the function $g(\xi)$
grows logarithmically with $\xi$ as $\xi \to \infty$, the  integral in 
(\ref{ggsigma})
is finite and gives~\cite{Ellis:1990hw}
\begin{eqnarray}
\sigma_{gg} &=& \frac{\as^4}{\pi m_Q^2} \frac{1}{N_c^2-1}
\left[ {23N_c^2\over 81} - {277\over 486} +
\left( {175 \zeta(3)\over 576} - {19\over 288} \right) {1\over N_c^2}
\right] \nn\\ &\approx& \frac{\as^4}{m_Q^2}\; 0.0803\, .\label{ggsigmasympt}
\end{eqnarray}
%%%%%%%%%%%%%%%%%%%%%%%%%%%%%%%%%%%%%%%%%%%%%%%%%%%%%%%%%%%%%%%%%%%%%%
\EPSFIGURE[ht]{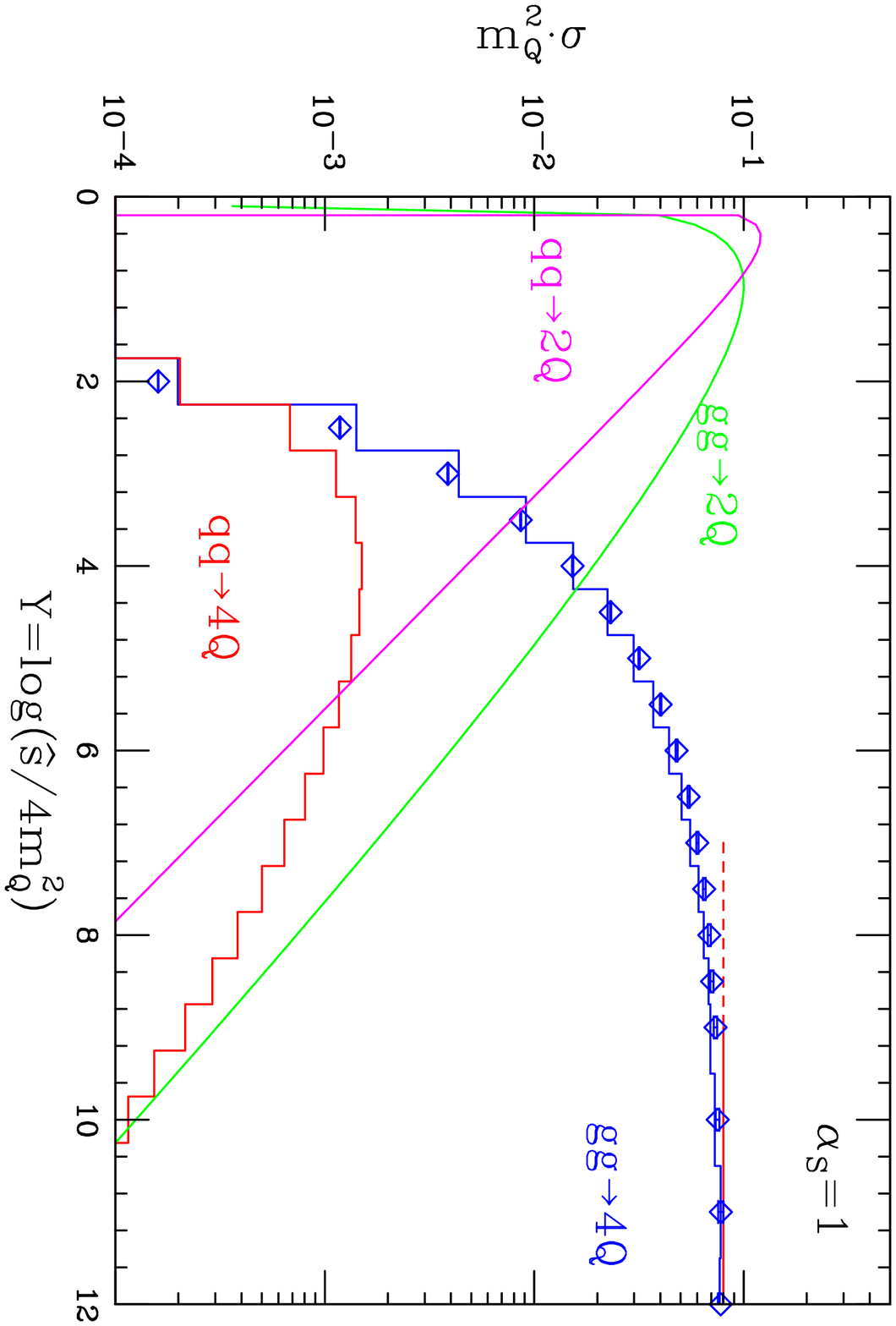,width=8cm,angle=90} {Partonic cross section for
$Q \Qb$ and $Q \Qb Q \Qb$ production.  The histograms show the exact
leading-order
results, i.e., the exact matrix elements integrated over the exact
phase space.  The diamonds are obtained integrating the high-energy
limit of the matrix element, Eq.~(\ref{ggQQQQfact}), with the exact
phase space.  The patterned red line is Eq.~(\ref{ggsigmasympt}),
representing the asymptotic limit. For comparison, the Born $gg,q\bar
q \to Q \bar Q$ contributions are also shown.  The coupling $\as$ is
set to one. Note that the kinematic limit for $b$-quark production 
at the Tevatron is at $Y\approx 10.6$.
\label{fig:partxsec} }
%%%%%%%%%%%%%%%%%%%%%%%%%%%%%%%%%%%%%%%%%%%%%%%%%%%%%%%%%%%%%%%%%%%%%%
The results obtained in this section are summarized in
Fig.~\ref{fig:partxsec}. The exact leading-order results for the
$gg\to Q\Qb Q\Qb$ and $q\bar{q}\to Q\Qb Q\Qb$ processes, obtained with
MADGRAPH/MADEVENT~\cite{Stelzer:1994ta,Maltoni:2002qb}, are shown
(histograms) as a function of $Y=\log(\hat{s}/4m_Q^2)$. The
dominance of the $t$-channel gluon exchange contribution, present only
in the case of $gg$ initial state, is apparent.  The diamonds are
obtained by integrating the high-energy limit of the matrix element,
Eq.~(\ref{ggQQQQfact}), with the exact phase space; the difference
with the exact result is fairly small, which implies that, at the
dynamical level, the high-energy limit is a good approximation. The
approximation of the phase space is evidently more drastic, and
results in the constant (dashed) red line, whose value is taken from
Eq.~(\ref{ggsigmasympt}).  For comparison we also show the Born
$gg,q\bar q \to Q \bar Q$ processes corresponding to the $a_{00}$
contribution in \eqn{BFKLxsec}. As argued above, and in contrast to
the $gg\to Q\Qb Q\Qb$ contribution, these exhibit a $\exp(-Y)$
behaviour in the high-energy (large $Y$) limit.

\section{The BFKL Monte Carlo}
\label{sec:bfklmc}

As we have seen, in the high-energy limit (\ref{hekin}) the cross section for the production
of four heavy quarks is dominated by processes with a gluon exchange in the
crossed channel. In that limit, the BFKL
formalism resums the universal
leading-logarithmic (LL) corrections, of ${\cal O}(\as^n\log^n(\hat{s}/
|\hat{t}|))$, with $\hat{t}$ defined in Eq.~(\ref{hattdef}). These are
obtained in the limit of strong rapidity ordering of the emitted gluon
radiation,
\begin{equation}
y_{Q_1}\simeq y_{\Qb_2} \gg y_1\gg y_2\gg \ldots \gg y_{n-1}\gg y_n
\gg y_{Q_3}\simeq y_{\Qb_4} \, ,\label{bfklkin}
\end{equation}
where we label by $1,\ldots, n$ the emission of $n$ gluons along
the BFKL ladder. Because of the strong rapidity ordering, the contribution
of the gluons to the parton momentum fractions (\ref{hemtmfr}) is
subleading, and it is therefore neglected to LL accuracy.
The BFKL-resummed cross section for the production of four heavy quarks
is then given by \eqn{xsec2}, where the $\delta$ function,
$\delta^2( {\bf q}_{a_\perp} - {\bf q}_{b_\perp} )/2$, is replaced
by the solution of the BFKL equation,
\begin{equation}
f({\bf q}_{a_\perp},{\bf q}_{b_\perp},\Delta y)\, =
{1\over (2\pi)^2 \sqrt{q_{a_\perp}^2 q_{b_\perp}^2} }
\sum_{n=-\infty}^{\infty} e^{in\phi}\, \int_{-\infty}^{\infty} d\nu\,
e^{\omega(\nu,n)\Delta y}\, \left(q_{a_\perp}^2\over q_{b_\perp}^2
\right)^{i\nu}\, ,\label{solc}
\end{equation}
with $\phi$ the azimuthal angle between $q_a$ and $q_b$,
and $\omega(\nu,n)$ the eigenvalue of
the BFKL equation with maximum at $\omega(0,0)=4\log{2}C_A\as/\pi$.
Thus the solution of the BFKL equation resums powers of $\Delta y$,
and rises with $\Delta y$ as $f({\bf q}_{a_\perp},{\bf q}_{b_\perp},\Delta y)
\sim\exp(\omega(0,0)\Delta y)$.

However, in a comparison with experimental data, it must be remembered
that the LL BFKL resummation makes some approximations which, even
though formally subleading, can be numerically important: $a)$ the
BFKL resummation is performed at fixed coupling constant, and thus any
variation in the scale at which $\as$ is evaluated appears in the
next-to-leading-logarithmic (NLL) terms; $b)$ because of the strong
rapidity ordering any two-parton invariant mass is large. Thus there
are no collinear divergences in the LL resummation in the BFKL ladder;
$c)$ finally, energy and longitudinal momentum are not conserved,
since the momentum fractions $x$ of the incoming partons are
reconstructed from the kinematic variables of the four heavy quarks
only, without including the radiation from the BFKL ladder. Therefore,
the BFKL theory will severely underestimate the correct value of the
$x$'s, and thus grossly overestimate the gluon luminosities. In fact,
if four heavy quarks $+\ n$ gluons are produced, the correct
evaluation of the $x$'s yields
\begin{equation}
x_a = \sum_{i=1}^4 {m_{Q_{i\perp}} e^{y_{Q_i}}\over \sqrt{S} }
+\ \sum_{j=1}^n {p_{j\perp}e^{y_i}\over\sqrt{S}} \qquad
x_b = \sum_{i=1}^4 {m_{Q_{i\perp}} e^{-y_{Q_i}}\over \sqrt{S} }
+\ \sum_{j=1}^n {p_{j\perp} e^{-y_i}\over\sqrt{S}}\,
,\label{bfklmtmfr}
\end{equation}
where $p_{j\perp}$ are the transverse momenta of the gluons produced
along the BFKL ladder.

In the standard (analytic) approach to BFKL, which leads to Eq.~(\ref{solc}),
it is not possible to take the contribution of the BFKL gluon radiation into
account in Eq.~(\ref{bfklmtmfr}). This is because in deriving
Eq.~(\ref{solc}) one has already integrated over the full rapidity ordered
phase space for BFKL gluon radiation. To gain information on the BFKL gluon
momenta we need to unfold the gluon integrations. This approach results in an
explicit sum over the number of emitted BFKL gluons, where each term in the
sum is an integral over the rapidity ordered BFKL gluon phase space. The
solution to the BFKL equation can then be obtained (numerically) while
maintaining information about each emitted gluon by evaluating these
integrals in a Monte Carlo approach~\cite{Schmidt:1996fg,Orr:1997im}. Besides
allowing energy and momentum conservation to be observed by including the
BFKL gluon contribution to Eq.~(\ref{bfklmtmfr}), this approach also allows
subleading effects originating from the running of the coupling to be taken
into account. The method has recently been generalised to solve the BFKL
equation at full NLL accuracy~\cite{Andersen:2003an,Andersen:2003wy},
although some work remains to be done before it can be applied in a
phenomenological study like the one presented here.

The Monte Carlo formulation of Ref.~\cite{Orr:1997im} is, in its simplest form,
applicable only when the transverse momentum of at least one end of the BFKL
chain is kept bigger than some cut-off $|\mathbf{q}_i|>P_{\perp}\gg\mu$ with
$i\in\{a,b\}$, and $\mu$ the resolution scale of the BFKL Monte Carlo (see
Ref.~\cite{Orr:1997im} for further details). It was demonstrated in
Ref.~\cite{Orr:1997im} that in the case of hadronic dijet production with a
minimum $P_\perp=20$~GeV, the residual $\mu$-dependence is negligible for $\mu\leq
6$~GeV. Varying $\mu$ will shift contributions between different $f^{(n)}$'s
describing the contribution from  different numbers of resolved gluons.

However, in the current process of $4Q$ production there is no minimum
transverse momentum scale at either end of the BFKL chain. To resolve the
problem thus faced by the BFKL MC formulation we cut out a small region of
phase space corresponding to $p_{\perp}<0.05$~GeV at one end of the
chain. The contribution from this very small region of phase space is
negligible, but nevertheless this cut-off is sufficient to permit the use of the unfolded
BFKL formalism. In principle $\mu$ could then be chosen arbitrarily small compared to the
cut-off, but this would result in very slow convergence due to the extremely
large number of resolved gluons with a transverse momentum above this scale.
Instead, $\mu$ is chosen according to the transverse momentum at one end of
the BFKL chain in 5 steps.  This keeps the average number of resolved gluons
under control and thus ensures rapid convergence, while maintaining the very
weak $\mu$-dependence of the overall result.

In order to demonstrate the behaviour of the BFKL ladder,
we consider the production of four heavy quarks assuming that all of them
are detected. We study the production rate as a function of the transverse momentum ${\bf
  q}_{a_\perp} = - {\bf p}_{Q_{1\perp}} - {\bf p}_{\Qb_{2\perp}}$ exiting
from the impact factor $I^{Q\Qb}(q_a)$. At leading order, the transverse momenta of the
two pairs are equal, ${\bf q}_{a_\perp} = {\bf q}_{b_\perp} = {\bf
  q}_{{}_\perp}$.  Since we know from \eqns{intif1}{intif2} that the
scaling of the integrated impact factor is $g(\xi) \sim
\ord(\xi)$, with $\xi= q_\perp^2/m_Q^2$, power counting from \eqn{ggsigma}
shows that at leading order $d\sigma/dq_{{}_\perp} \sim \ord(q_{{}_\perp})$ as
$q_{{}_\perp}\to 0$.  When the BFKL gluon radiation is included%
%% through
%% \eqn{solc}
, the production rate is hardened in the infrared and we obtain
\begin{equation}
{d\sigma\over dq_{a_\perp} dq_{b_\perp} } \sim {\rm const.} \qquad
{\rm as} \qquad  q_{a_\perp}\to 0\ ,\quad q_{b_\perp}\to 0\,.
\end{equation}
In \fig{fig:ifpt} we plot the transverse momentum distribution $d\sigma/
dq_{a_\perp} dq_{b_\perp}$ evaluated at $q_{a_\perp}=q_{b_\perp}=p_{\perp}$.
The solid red curve is the four-quark production (\ref{xsec2}), but with the
high-energy parton momentum fractions replaced by the exact ones, $x_a^0\to
x_a$ and $x_b^0\to x_b$; in this case, the two impact factors have equal
transverse momenta ${\bf q}_{a_\perp} = {\bf q}_{b_\perp}$. The dashed blue
curve corresponds to the high-energy limit of leading-order 
four-quark production (\ref{xsec2})
with the BFKL ladder included. In this case ${\bf q}_{a_\perp}$ is no longer
restricted to be equal to ${\bf q}_{b_\perp}$, which explains why the BFKL
curve is lower than the leading-order one. However, we see that the spectrum is
relatively harder for $p_\perp\to 0$ in the BFKL case.

\EPSFIGURE[ht]{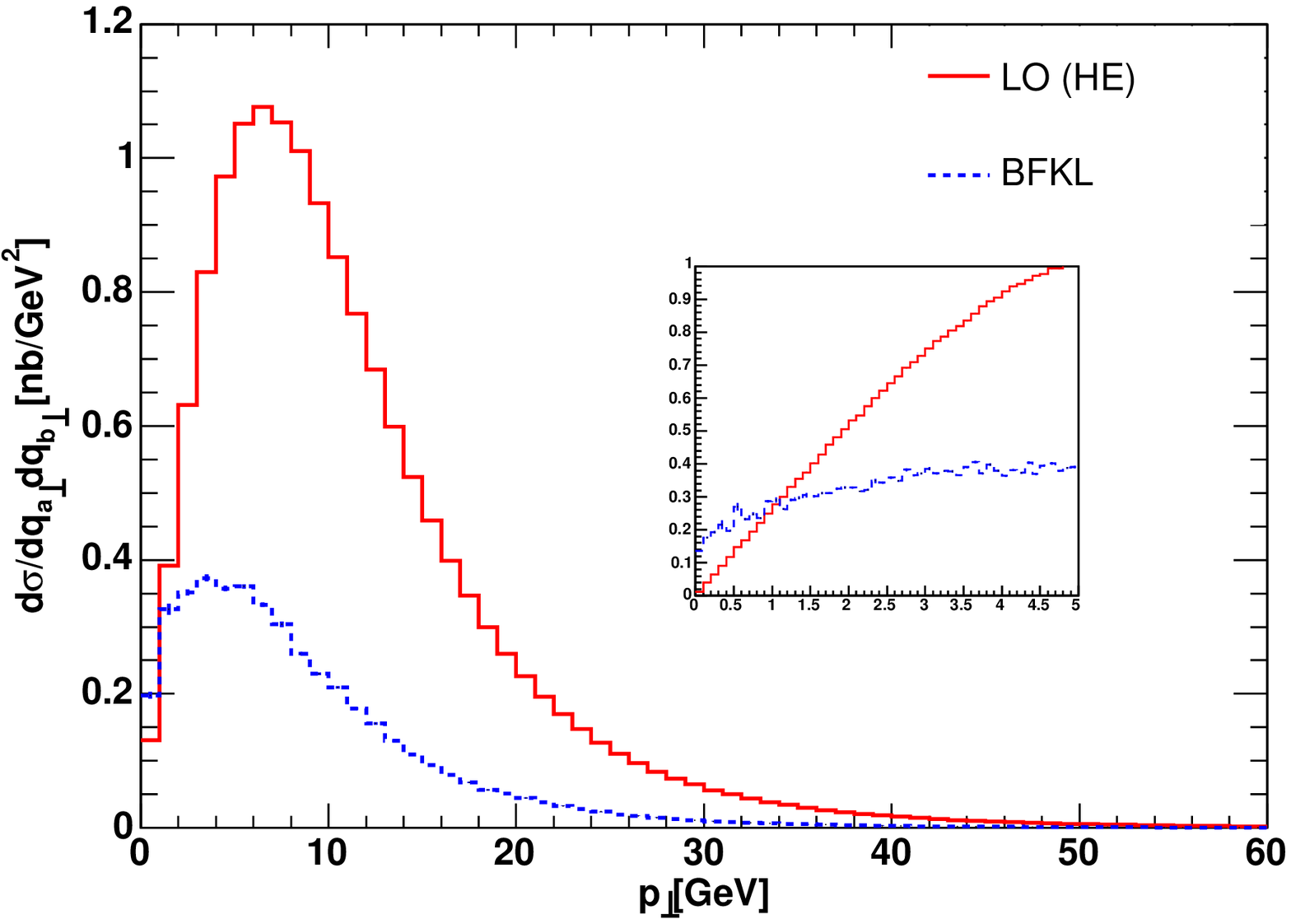,width=10cm}
{The transverse momentum distribution $d\sigma/ dq_{a_\perp}dq_{b_\perp}$
  evaluated at $dq_{b_\perp}\!=\!dq_{b_\perp}\!=\!p_\perp$.
The solid red curve corresponds to the high-energy limit of 
leading-order four $b$-quark
production, with $m_b = 5$~GeV. The dashed blue curve
corresponds to adding BFKL evolution to the gluon exchanged in the $t$-channel.
\label{fig:ifpt} }

\section{BFKL signals at the Tevatron and LHC}
\label{sec:bfklobs}

\subsection{Inclusive heavy-quark production}
\label{sec:bfklobs2q}

In this section we compare the results for the $4Q$ channel, obtained
with the BFKL MC described in the previous section, with those relevant to 
$Q\Qb$ production, obtained with the NLO code of Ref.~\cite{Mangano:jk}
and MC@NLO~\cite{Frixione:2002ik,Frixione:2003ei}.
We consider bottom quark production, with $m_b=5$~GeV, since $b$-quarks are
readily identifiable at the Tevatron and LHC. In the case of
pair production, we need to use a NLO computation in order to explicitly
verify that, with our chosen set of cuts, large non-BFKL logarithms do not
appear in the cross section, which is a necessary condition in order
to study BFKL signals with the $4Q$ channel.

\begin{figure}[t]
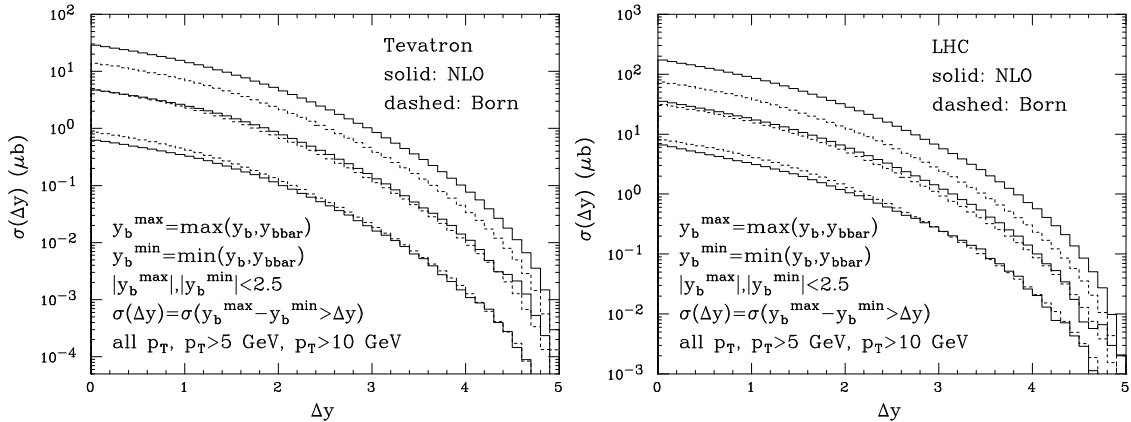

  \begin{center}
    \epsfig{figure=bbdsdy_tev.ps,width=0.49\textwidth}
    \epsfig{figure=bbdsdy_lhc.ps,width=0.49\textwidth}
\caption{\label{fig:bbdsdy}
Integrated cross sections as a function of $\Delta y$, at LO (dashed
histograms) and NLO (solid histograms), with no cut on the transverse
momentum $p_{\rm T}$, and with $p_{\rm T} > 5$ and 10~GeV, at the
Tevatron (left panel) and LHC (right panel) energies.  The code of
Ref.~\cite{Mangano:jk} has been used.
}
  \end{center}
\end{figure}

Figure~\ref{fig:bbdsdy} shows the integrated cross section
\begin{equation}
\sigma(\Delta y)=\int_{\Delta y}^\infty d\Delta y^\prime
\frac{d\sigma}{d\Delta y}(\Delta y^\prime)
\end{equation}
as a function of $\Delta y$, the rapidity distance between the two
tagged quarks (which, for this process, are $b$ and $\bar{b}$), at
Tevatron and at LHC energies.  In order to simulate a realistic
detector coverage, the rapidity of both quarks is required to be less
than 2.5, and therefore $\Delta y=5$ is the largest accessible
rapidity separation. We also consider additional cuts on the
transverse momenta of the tagged quarks, imposing $p_{{\rm T} b,\bar
b} > 5$ and 10~GeV.  The two-loop running of the strong coupling
$\as$, and the MRST99 package~\cite{Martin:1999ww} of pdfs has been
used, with factorisation scale set to $\mu_{\sss F}^2 =
(m_{b_{\perp}}^2 + m_{\bar{b}_{\perp}}^2)/2$.
%%%%%%%%%%%%%%%%%%%%%%%%%%%%%%%%%%%%%%%%%%%%%%%%%%%%%%%%%%%%%%%%%%%
%%%%%%%%%%%%%%%%%%%%%%%%%%%%%%%%%%%%%%%%%%%%%%%%%%%%%%%%%%%%%%%%%%%
>From the figure we can see that the cuts on the transverse momenta
largely reduce the impact of radiative corrections. 
However, this
information alone is not sufficient to guarantee that non-BFKL logs
do not spoil the perturbative expansion. In order to investigate
this issue, we thus recomputed the cross section with 
MC@NLO~\cite{Frixione:2002ik,Frixione:2003ei}, which, by matching
the NLO results with the {\small HERWIG}~\cite{Corcella:2000bw}
parton shower, improves the fixed-order result by effectively 
resumming various classes of large logs. In the case in which 
no $p_{\rm T}$ cuts are applied, the MC@NLO results are basically
coincident with the NLO ones. However, by imposing 
$p_{{\rm T} b,\bar b}>5$~GeV, the MC@NLO cross section is roughly
a factor 1.7 larger than the NLO, in the whole $\Delta y$ range
considered. This is due to the fact that the $p_{\rm T}$ cuts render the
cross section sensitive to Sudakov effects. Although these could be
reduced by imposing different $p_{\rm T}$ cuts on the two tagged $b$'s,
it is quite problematic to eliminate them completely. Thus, the
pure NLO result must be regarded, at least for the $p_{\rm T}$ cuts
considered here, as a lower bound on the $b\bar b$ inclusive
cross section.

\begin{figure}[t]
  \begin{center}
    \epsfig{figure=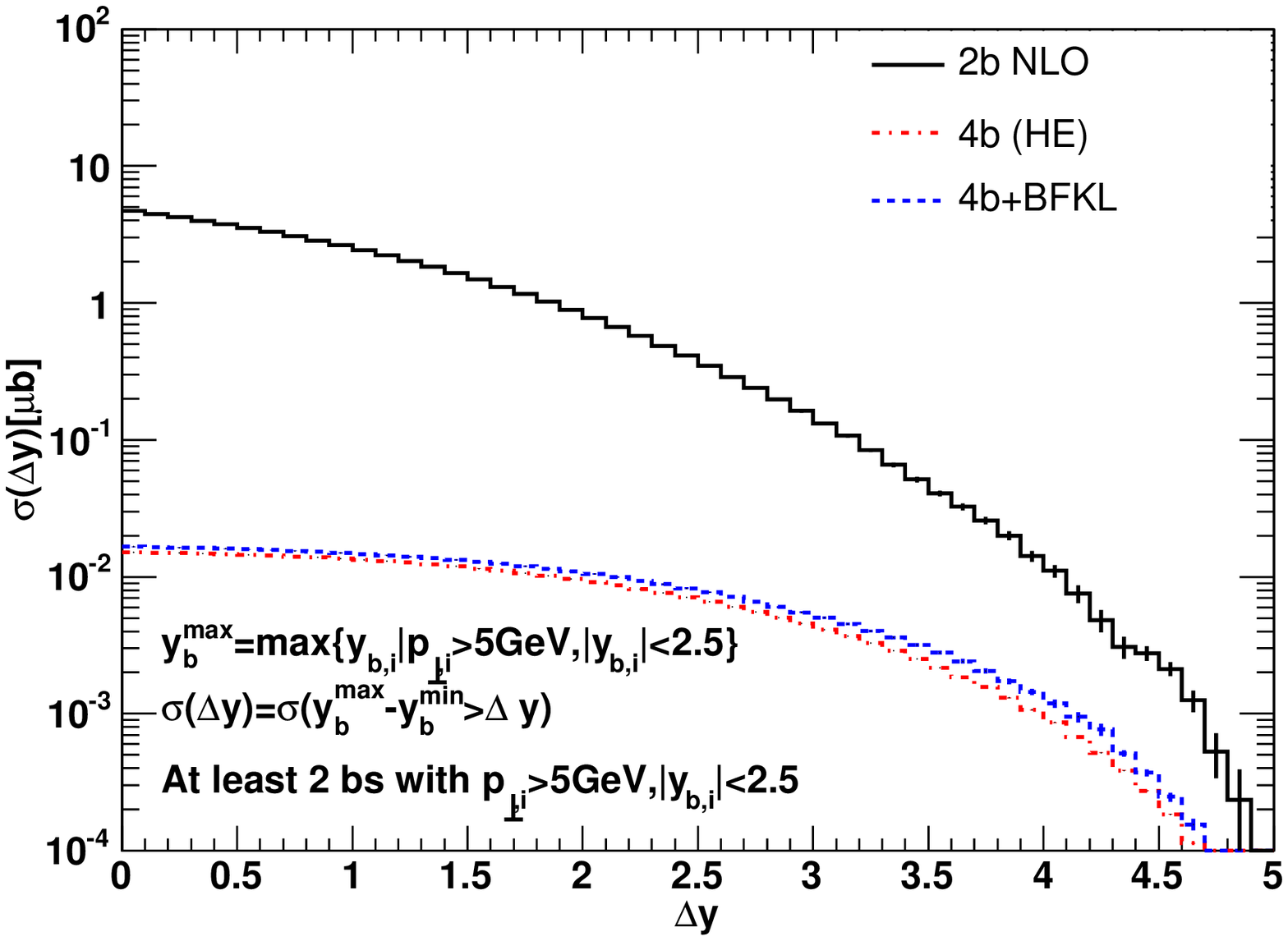,width=0.49\textwidth}
    \epsfig{figure=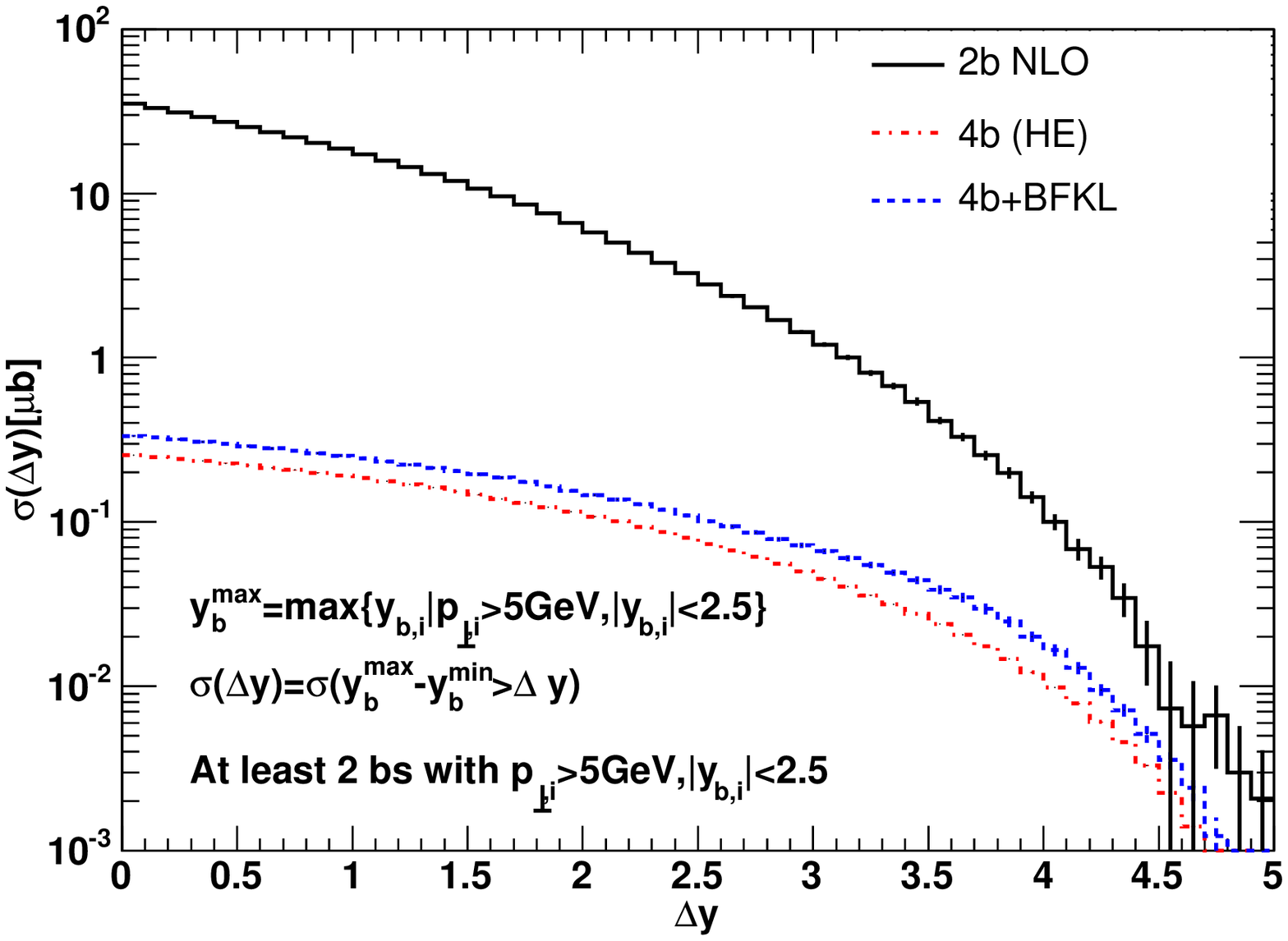,    width=0.49\textwidth}
\caption{\label{fig:bbbb}
Integrated cross sections as a function of $\Delta y$, with $p_{\rm T}> 5$ GeV, 
at Tevatron (left panel) and LHC (right panel) energies. The
NLO curves are the same as the middle NLO curves of Fig.~4,
and are displayed here for the sake of comparison. The dot-dashed red curves are
the high-energy limit contributions of the $4Q$ channel to inclusive
$2Q$ production; the dashed blue curves are the same as the red curves with
the addition of BFKL gluon radiation.}
  \end{center}
\end{figure}

In order to be definite, we require $p_{{\rm T} b,\bar b} > 5$~GeV
in what follows. In \fig{fig:bbbb}, we plot the integrated cross
section for $b\bar b$ production as a function of $\Delta y$, at
Tevatron and at LHC energies. For the sake of comparison, we
display again here the middle NLO curves of
\fig{fig:bbdsdy}. In addition, the dot-dashed red curve displays the high-energy
limit contribution of the $4Q$ channel to inclusive $2Q$ production,
where $Q = b $ or $\bar b$.  The factorisation and renormalisation
scales have been set to $\mu_{{\sss F}a}^2 = \mu_{{\sss R}a}^2 =
(m_{b_{1\perp}}^2 + m_{\bar{b}_{2\perp}}^2)/2$ and $\mu_{{\sss F}b}^2
= \mu_{{\sss R}b}^2 = (m_{b_{3\perp}}^2 + m_{\bar{b}_{4\perp}}^2)/2$.
Thus, the strong coupling $\as^4$ must be understood here as
$\as^2(\mu_{{\sss R}a}^2)\, \as^2(\mu_{{\sss R}b}^2)$, with $\as$
evolved at two loops, in accordance with the NLO
calculation.\footnote{ We justify the scale choices as follows: in the
high-energy limit the impact factors for $b\bar b$ production on
either side can be viewed as two almost independent scattering centres
linked by a gluon exchanged in the crossed channel. It therefore makes
sense to run the pdfs and $\as$ according to the scales set by each
impact factor.}  The dashed blue curve is the same as the red curve but
with the addition of BFKL gluon radiation. In the BFKL gluon emission
chain, the value of $\as$ is taken at the $b$ mass.

\fig{fig:bbbb} is the central result of this study. It shows that
within the rapidity range for heavy-quark production accessible to LHC
(assumed here to correspond to $\Delta y < 5$) the $4Q$ channel, even
augmented by the BFKL gluon radiation, can never overcome the $2Q$
channel.  Thus, it cannot readily be used as a footprint of BFKL
radiation.

The situation could be improved either by imposing the
additional requirement that the two tagged $b$-quarks have the same
sign, i.e. $bb$ or $\bar b \bar b$, or by requiring three or more
$b$-quarks to be identified.  In the former case, this would reduce
the ``$4Q$" curves in \fig{fig:bbbb} by a (combinatoric) factor of two,
while almost completely removing the $2b$ contribution.  
A realistic assessment of how much of the BFKL signal
would remain in these cases would depend on the efficiencies of
multi-$b$-quark tagging and charge identification (via the sign of the
lepton in semi-leptonic $B$-meson decay, for example), which goes
beyond the scope of the present study. Another issue that needs to be
addressed by a more realistic study is the contamination from
overlapping events, see for example Ref.~\cite{Mangano:2002wn}.
%%%%%%%%%%%%%%%%%%%%%%%%%%%%%%%%%%%%%%%%%%%%%%%%%%%%%%%%%%%%%%%%%%%
%%%%%%%%%%%%%%%%%%%%%%%%%%%%%%%%%%%%%%%%%%%%%%%%%%%%%%%%%%%%%%%%%%%
\subsection{Inclusive heavy-quark $+$ 1 jet production}
\label{sec:bfklobs2q1j}

As mentioned in the Introduction and in \sec{sec:hel}, inclusive
$Q\bar{Q} + 1$~jet production is also of interest in the high-energy limit,
and is in a sense a hybrid of the original Mueller-Navelet 2 jet and 
our $4Q$ processes.
In this process, a gluon is exchanged in the $t$-channel between the jet 
and the $Q\bar{Q}$ pair already at leading-order, which in this case is
$\ord(\as^3)$. In fact, as in \eqn{BFKLxsec}, the $\Delta y$ distribution 
for $Q\bar{Q} + 1$~jet production in the high-energy limit can be written 
schematically as
\beq
\frac{d\sigma_{Q\bar Q jet}}{\Delta y}
\,\sim\, \as^3 \sum_{j=0}^\infty b_{0j} \as^j\
+\  \as^3 \sum_{j=0}^\infty b_{1j} (\as L)^j\ +\ \as^3 \sum_{j=0}^\infty
b_{2j} \as(\as L)^j\ + \cdots,
\label{BFKLxsec1}
\eeq
where \mbox{$L=\log(\hat s/\mu_{\sss\rm W}^2) \simeq\Delta y$} 
is a large logarithm, and the quantity $\mu_{\sss\rm W}^2$ is a
mass scale squared. The first sum in \eqn{BFKLxsec1}
is a fixed-order expansion in $\as$ starting at $\ord(\as^3)$, and
collects the contributions which do not feature gluon exchange 
between the jet and the $Q\bar{Q}$ pair.
The $b_{00}$ coefficient is the leading-order term (a specimen diagram is 
depicted in \fig{fig:bb}(b)). The second and third sums of \eqn{BFKLxsec1}
collect the contributions which feature only gluon exchange in the crossed
channel between the jet and the $Q\bar{Q}$ pair, the second (third) sum 
resumming the BFKL (next-to-)~leading logarithmic corrections. 
\fig{fig:bb}(c) represents the zeroth-order term of the second sum. 
The $b_{1j}$ and $b_{2j}$ coefficients behave like $1/\mu_{\sss\rm W}^2$.
We note, however, that in contrast to \eqn{BFKLxsec},
the second and third sums of \eqn{BFKLxsec1} start
at the same order in $\as$ as the first sum. Thus one would expect that
the onset of the dominance of the asymptotic energy region $\hat{s}\to\infty$
occurs more quickly in this case than in heavy two-quark production.
We analyse this issue by computing the coefficients $b_{00}$ and $b_{1j}$.

\EPSFIGURE[ht]{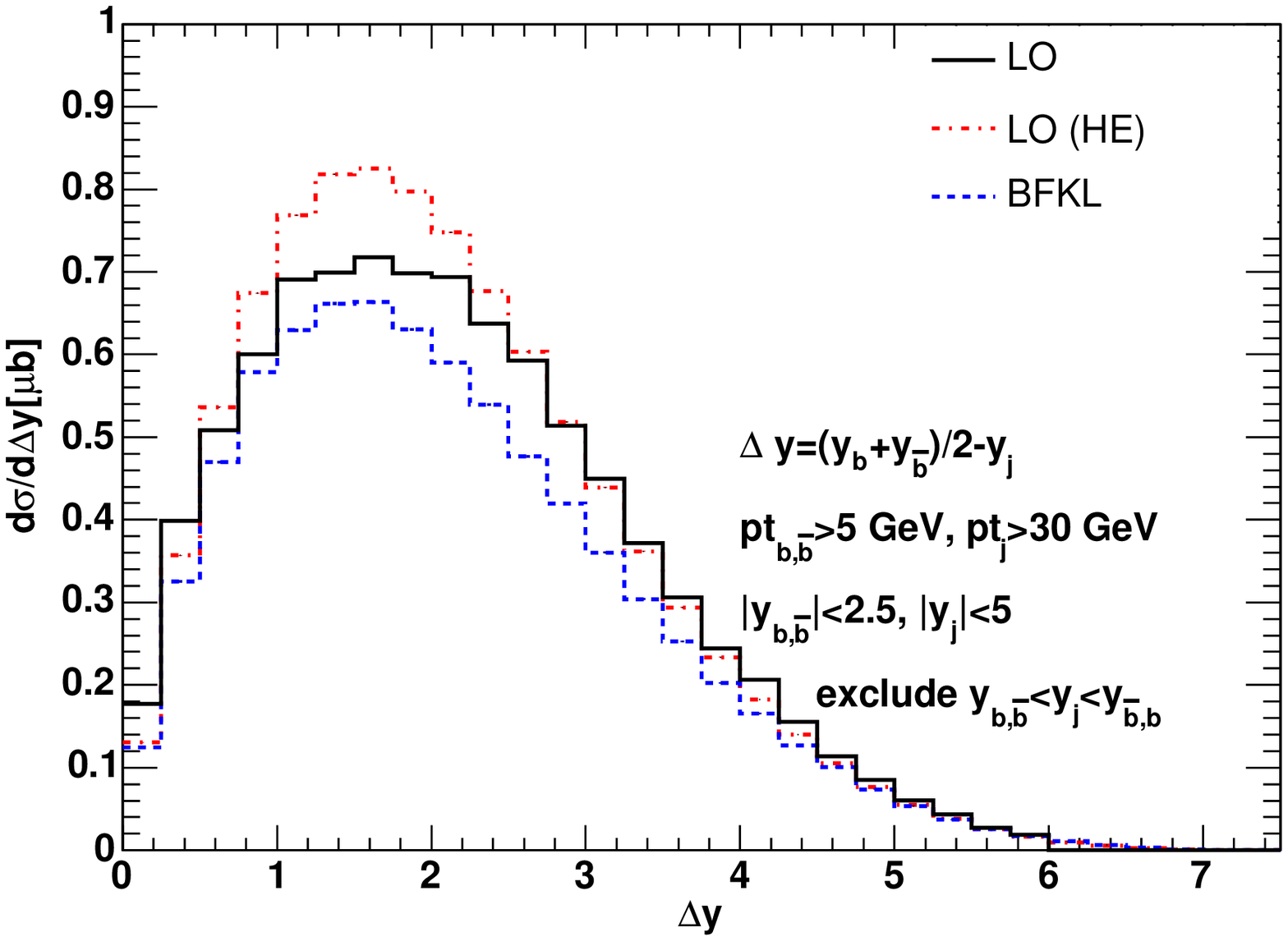,width=12cm}
{Inclusive heavy two-quark $+ 1$~jet production
as a function of the rapidity separation of the jet from the average position
of the heavy two-quark pair, $\Delta y = |y_j - (y_b+y_{\bar b})/2|$. 
The solid and red dot-dashed curves correspond to leading-order production,
exactly and in the high-energy limit respectively.
The dashed blue curve is the leading order plus BFKL resummation.
\label{fig:bbjet} }

We consider inclusive heavy two-quark $+\ 1$~jet production in the high-energy
limit. The heavy quarks are $b$ quarks for which, following the analysis of
\sec{sec:bfklobs2q}, we require that $|y_{b,\bar b}| < 2.5$ and 
$p_{{\rm T} b,\bar b} > 5$~GeV.
For the jet, we require the set of cuts $|y_j| < 5$ and 
$p_{{\rm T} j} > 30$~GeV.
The factorisation and renormalisation scales are taken as 
 $\mu_{{\sss F}a}^2 = \mu_{{\sss R}a}^2 = 
(m_{b_{1\perp}}^2 + m_{\bar{b}_{2\perp}}^2)/2$
and $\mu_{{\sss F}b}^2 = \mu_{{\sss R}b}^2 = p_{{\rm T} j}^2$, since
the impact factors for $b\bar b$ production on one side and
for jet production on the other can be viewed as two almost independent
scattering centres linked by a gluon exchanged in the crossed channel.
Thus the strong coupling $\as^3$ must be understood here as 
$\as^2(\mu_{{\sss R}a}^2)\, \as(\mu_{{\sss R}b}^2)$.
In \fig{fig:bbjet}, we show the distributions for heavy two-quark $+\ 1$~jet 
as a function of the rapidity separation of the jet from the average 
position of the heavy two-quark pair, $\Delta y = |y_j - (y_b+y_{\bar b})/2|$.
The solid curve corresponds to leading-order production (exact matrix element); 
the dot-dashed red curve is the leading-order production in the high-energy limit
approximation and the dashed blue curve is the leading order plus BFKL resummation, 
as given by the  Monte Carlo generation of the ladder gluons 
(i.e.\ with energy-momentum conservation).

There is evidently a sizeable {\em suppression} from the resummation when the BFKL gluons are 
radiated off the ladder while conserving energy-momentum, reminiscent
of what happens in the case of dijet production in the high-energy limit
\cite{Orr:1998hc}. This is at first sight puzzling, because the kinematics of two-quark 
$+\ 1$~jet production in the high-energy limit resemble more closely the ones of
$W +\ 2$~jet production rather than those of dijet production, and in
$W +\ 2$~jet production in the high-energy limit there is no such strong
suppression when enforcing energy-momentum conservation on the BFKL ladder
\cite{Andersen:2001ja}. However, that is where the similarity ends: in
$W +\ 2$~jet production, the impact factor for $W +\ 1$~jet production 
is generated by a {\em quark}, while in the present case the impact factor for
$b\bar b$ production is generated by a {\em gluon}, and therefore the dependence
on the pdfs in the two cases is completely different.\footnote{In order to
rule out other possible explanations, we tried to mimic as much as
possible the set-up of $W +\ 2$~jet in two-quark $+\ 1$~jet production, 
namely we eliminated the gluon-gluon sub-process, so as to make two-quark 
$+\ 1$~jet production by quark-gluon scattering the dominant process, and we set the $b$-quark mass
equal to the $W$ mass. Even with these modifications,  we still obtain a BFKL distribution
 with same qualitative features as in \fig{fig:bbjet}.}

\section{Conclusions}
\label{sec:conc}

A definitive test of BFKL physics at hadron colliders is still
lacking. A number of processes have been suggested, including the
standard Mueller-Navelet dijet production, and in this paper we have
studied a new possibility: four heavy-quark production with a large
rapidity separation between two of the heavy quarks. The common
feature of all these `BFKL' processes is the presence of a $t$-channel
gluon in the scattering amplitude, which gives the dominant
contribution in the high-energy limit.

In this work we have focused on the production of $b$ quarks at Tevatron and LHC
energies. The simplest quantity to measure is the 2$b$ inclusive cross
section as a function of the rapidity separation $\Delta y$. However
in this case the $4b$ process has to compete with leading- and
next-to-leading-order $b \bar b$ production. Using a set of
representative cuts on rapidities and transverse momenta, we have
shown that in practice the NLO $b \bar b$ contribution is dominant
over the measurable $\Delta y$ range, although at the very highest
$\Delta y$ values ($\sim 5$) at the LHC energy the $b \bar b$ and $4b$
contributions are of comparable magnitude.

We can conclude, therefore, that it will be very difficult to detect
any BFKL signal in the $2b$ inclusive distribution. However, a
characteristic feature of the $4Q$ contribution in the high-energy
limit is that the two heavy quarks separated by a large rapidity
distance are as likely to have the same as opposite sign.  The ability
to tag the sign of the $b$ quarks could therefore be used eliminate
the NLO $b \bar b$ contribution.  We note also that in the case of the
$4b$ process, the bottom quantum number is conserved locally in
rapidity, i.e. many of the events with two detected $b$ quarks with a large
rapidity separation could have one or two additional $b$ quarks in the
detector. To study these possibilities in detail would however require
detailed knowledge of the detector capability, and is therefore beyond
the scope of the present work.

Finally, we also considered the case of $Q\bar Q$ + 1~jet production,
which is an extension of the original dijet case in which one of the
far forward/backward jets is replaced by an heavy-quark pair. Here
there is a $t$-channel gluon already at leading order and so one might
expect an earlier onset of the high-energy asymptotic regime. However,
because the dominant contribution involves gluons in the initial
state, there is a severe suppression from the pdfs when the additional
energy radiated in the BFKL ladder is properly taken into account.
This means that for this process, the fixed-order perturbative
contribution (i.e. LO or NLO) is likely to be a good approximation to
the full cross section over the accessible kinematic range.

\section*{Acknowledgments}

Useful discussions with Albert De Roeck, Keith Ellis, Michelangelo
Mangano and Stefan Tapprogge are gratefully acknowledged.  The authors
would like to thank the CERN Theory Division for hospitality which
enabled this work to be completed. VDD thanks the IPPP at the
University of Durham for similar hospitality. This research is
supported in part by the UK Particle Physics and Astronomy Research
Council.

\appendix

\section{Four heavy-quark production}
\label{sec:4heavy}

The partonic cross section for four heavy-quark production is
\begin{equation}
d{\hat \sigma}(p_a p_b\to  p_{Q_1} p_{\Qb_2} p_{Q_3} p_{\Qb_4})
= {1\over 2{\hat s}}
d{\cal P}_4 |{\cal M}_{p_a\, p_b \to Q_1 \Qb_2 Q_3 \Qb_4 }|^2\,
,\label{genps}
\end{equation}
with $\hat s = (p_a+p_b)^2 = x_ax_b S$ and $S$ the squared partonic
and hadronic centre-of-mass energies respectively,
and with four heavy-quark phase space
\begin{equation}
d{\cal P}_4 = \prod_{i=1}^4 {d^3 p_{Q_i}\over (2\pi)^3 2p^0_{Q_i}}\,
(2\pi)^4 \, \delta^4(p_a + p_b - p_{Q_1} - p_{\Qb_2} - p_{Q_3} - p_{\Qb_4})\,
,\label{hadphase}
\end{equation}
with $p^0_{Q_i} = \sqrt{{\bf p}_{Q_i}^2 + m_Q^2}$.
The factorisation formula is
\begin{equation}
d\sigma = \sum_{ab}\,dx_a dx_b\ f_{a/A}(x_a,\mu_F^2)\,
f_{b/B}(x_b,\mu_F^2)\, d\hat\sigma\, ,\label{xfact}
\end{equation}
where the sum is over the parton species,
and $f_{a/A}(x_{a},\mu_F^2)$ is the pdf of the parton
$a$ of momentum fraction $x_{a}$ within hadron $A$, and similarly for
parton $b$.
Parametrizing the heavy-quark momenta in terms of the rapidities,
\begin{equation}
p_{Q_i} = (m_{Q_{i\perp}} \cosh{y_{Q_i}}, {\bf p}_{Q_{i\perp}},
m_{Q_{i\perp}} \sinh{y_{Q_i}})\, ,
\end{equation}
we can write the cross section for heavy-quark production as
\begin{eqnarray}
\lefteqn{
{d\sigma\over \prod_{i=1}^4 d^2{\bf p}_{Q_{i\perp}} dy_{Q_i} } }
\label{xsec}\\ &=&
\sum_{ab}\, {f_{a/A}(x_a,\mu_F^2)\, f_{b/B}(x_b,\mu_F^2)\over
x_ax_b}
{|{\cal M}_{ij}|^2\over (2\pi)^4 (4\pi)^4 S^2}\,
\delta^2( {\bf p}_{Q_{1\perp}} + {\bf p}_{\Qb_{2\perp}} +
{\bf p}_{Q_{3\perp}} + {\bf p}_{\Qb_{4\perp}})\, ,\nn
\end{eqnarray}
with momentum fractions of the incoming partons given by
\begin{equation}
x_a = \sum_{i=1}^4 {m_{Q_{i\perp}} e^{y_{Q_i}}\over \sqrt{S} }\; , \qquad
x_b = \sum_{i=1}^4 {m_{Q_{i\perp}} e^{-y_{Q_i}}\over \sqrt{S} }\,
.\label{mtmfr}
\end{equation}

\section{Impact factor for $g g^* \to Q \Qb$}
\label{sec:ifqq}

The impact factor, $I^{Q\Qb}$, for $Q \Qb$ production
can be obtained by using the squared
amplitude for $g\, i\to Q \Qb  i$ with $i=q,g$  from
Refs.~\cite{Ellis:1990hw,Mangano:jk}.
The momenta of the incoming and outgoing partons are
$g(p_a)+i(p_b)= Q(p_Q)+ \Qb(p_{\Qb} )+i( p_{b'} )$.
In the high-energy limit, the rapidities are strongly ordered while
the transverse momenta are of similar size,
\begin{equation}
y_Q\simeq y_{\Qb} \gg y_i\, ,\qquad p_{Q\perp} \simeq p_{\Qb\perp} \simeq
p_{i\perp} \; .
\end{equation}
The squared amplitude for $g\, i\to Q \Qb  i$,
summed (averaged) over final (initial) colours and helicities,
then reduces to
\begin{equation}
|{\cal M}_{g\, i\to Q\Qb\, i}|^2 =
\frac{4 \hat{s} ^2}{\hat{t} ^2} I^{Q\Qb}(p_a,p_Q,p_{\Qb};q)
I^{i}(p_b, p_{b'})\, ,\qquad i=q,g\, ,\label{QQfact}
\end{equation}
with $\hat s = (p_a+p_b)^2$ the squared centre-of-mass energy, and
$\hat t = (p_b-p_{b'})^2$ the momentum transfer.
The impact factor $I^{i}(p_b, p_{b'})$ for quark/gluon production,
summed (averaged) over final (initial) helicities and colours,
can be written as~\cite{Andersen:2001ja}
\begin{equation}
I^{g} = g^2\, {C_A\over N_c^2-1}\, \delta^{cc'}\, ,\qquad
I^{q} = {g^2\over 2N_c}\, \delta^{cc'}\, ,\label{impgg}
\end{equation}
where $C_A=N_c=3$, the index $c$ runs over the colours of the
gluon exchanged in the
crossed channel, and we have used the standard
normalization of the SU($N_c$) matrices, ${\rm tr} (\lambda^c \lambda^{c'}) =
\delta^{cc'} /2$. The impact factor for $g g^* \to Q \Qb$,
summed (averaged) over final (initial) colours and helicities, is then
\begin{eqnarray}
I^{Q\Qb}(p_a,p_Q,p_{\Qb};q)=&&
 \frac{\gs^4\; \delta^{cc'}}{4 N_c (N_c^2-1) \, t'_{a\Qb}\,  t'_{aQ} }
    \left[ \hat{t}\,
     \left( 1 + 2 N_c^2 \,\frac{t'_{a\Qb}}{s_{Q\Qb}}\,x \right) \,
     \left( x^2 + \tilde{x}^2 \right) \right. \nonumber\\
    && +\left.\frac{ 4 \, m_Q^2}{t'_{a\Qb}}\;
     \left(
     N_c^2 \frac{\,t'_{aQ} }{s_{Q\Qb}^2}
     \left( t'_{a\Qb}+ x s_{Q\Qb}\right)^2 +\,
      x \left (\tilde{x} t'_{a\Qb}- x  t'_{aQ} \right)
      \right)
      \right]
\nonumber\\
&&+( Q \leftrightarrow \Qb, x \leftrightarrow \tilde{x})\,,
\label{QQimp}
\end{eqnarray}
where we have defined the momentum fraction
\begin{equation}
x = \frac{p^+_Q}{p^+_Q + p_{\Qb}^+ }=1-\tilde{x}\, ,
\end{equation}
and the invariants
\begin{eqnarray}
&&s_{Q\Qb}  = (p_Q+p_{\Qb} )^2         \stackrel{{\rm h.e.}}{=}
\hat{t}+\frac{m_{Q\perp}^2}{x}+\frac{m_{\Qb\perp}^2}{\tilde{x}}\,,\nn\\
&&t'_{aQ}   = (p_a-p_Q)^2-m_Q^2        \stackrel{{\rm h.e.}}{=}
-\frac{m_{Q\perp}^2}{x}\,,\label{HErelations}\\
&&t'_{a\Qb} = (p_a-p_{\Qb} )^2-m_Q^2   \stackrel{{\rm h.e.}}{=}
-\frac{m_{\Qb\perp}^2}{\tilde{x}}\,.\nn
\end{eqnarray}
In the small $q_\perp$ limit, the jet opposite to the impact factor
for $Q\Qb$ production becomes collinear, and the cross section obtained
from the squared amplitude
(\ref{QQfact}) yields an infrared singular real correction.
Since the latter may have at most a logarithmic enhancement as
$q_\perp\to 0$, the squared amplitude (\ref{QQfact}) cannot diverge more
rapidly than $1/q_\perp^2$. This means that in the small $q_\perp$
limit, the impact factor  must be at least quadratic in
$q_\perp$, $I^{Q \Qb} \sim \ord(q_\perp^2)$. Using
${\bf q}_\perp = - ({\bf p}_{Q\perp} + {\bf p}_{\Qb_\perp})$,
we see immediately that this is the case.
In addition, as $q_\perp\to 0$ we have an almost on-shell gluon scattering
with a gluon, then ${\bf p}_{Q\perp}\to - {\bf p}_{\Qb\perp}$ and averaging
over the azimuthal angle of ${\bf q}_\perp$, Eq.~(\ref{QQimp}) becomes
\begin{eqnarray}
\lim_{q_\perp\to 0} I^{Q \Qb}
&=&\delta^{cc'}  \frac{\gs^4}{N_c (N_c^2-1)}
\frac{q_\perp^2\, x \tilde x\, } { m_{Q\perp}^4  }\,
\frac{(N_c^2 -1) - 2 N_c^2 x \tilde x}{2}
\left[1- 2 x \tilde x \left(1-
2 m_Q^2   \frac{p_{Q\perp}^2}{m_{Q\perp}^4}
\right) \right] \nonumber \\
&=&  \delta^{cc'} \left(\frac{q_\perp\, x  \tilde x } {
m_{Q\perp}^2 } \right)^2  |{\cal M}_{g\, g\to
Q\Qb}|^2 \, ,\label{QQimp3}
\end{eqnarray}
where for the invariants in the
$g\, g\to Q\Qb$ Born amplitude we have used
\begin{eqnarray}
&&s=\frac{m_Q^2+p^2_{Q\perp}}{x \,\tilde{x}}\,,\nn\\
&&t=-\frac{x m_Q^2+p^2_{Q\perp}}{\tilde{x}}\,,\\
&&u=2 m_Q^2-t -s\,.\nn
\end{eqnarray}
Equation~(\ref{QQimp3})
explicitly shows that in this limit the impact factor is positive definite,
and that it factorises into the squared amplitude
for $g\, g\to Q\Qb$ scattering.

\subsection{The integrated impact factor for $g g^* \to Q \Qb$}
\label{sec:appa1}

Using \eqn{QQimp} and the invariants (\ref{HErelations}), the
integrated impact factor (\ref{intif}) becomes
\begin{eqnarray}
{\cal I}(\xi)&=& {\as^2 \delta^{cc'}\over 2 N_c (N_c^2-1)}
\int_0^1 dx\;\int \frac{d^2 {\rm p}}{\pi}\;
\\ && \left [ (x^2+\tilde{x}^2)\, {\rm k}^2\;
\left(-\frac{1} {D_1 D_2}
  + \frac{N_c^2 x^2} {D_1 D_3}
  + \frac{N_c^2 \tilde{x}^2} {D_2 D_3}\right) \right. \nonumber\\
&&+ 4 \,x\, \tilde{x}\, m_Q^2 \;\left.
\left(\frac{1}{D_1 D_2}-\frac{N_c^2}{D_1 D_3}-\frac{N_c^2}{D_2 D_3}
+\frac{N_c^2-1}{2 D_1^2}+\frac{N_c^2-1}{2 D_2^2}+\frac{N_c^2}{D_3^2} \right)
\right]\, ,\label{intif0}
\end{eqnarray}
where on the left-hand side we have made explicit that the impact factor
depends only on the dimensionless ratio $\xi= q_\perp^2/m_Q^2$.
In \eqn{intif0} the propagators
\begin{eqnarray}
&&D_1=m_Q^2+ p^2_\perp \nonumber \\
&&D_2=m_Q^2+({\bf p}_\perp + {\bf k}_\perp)^2 \\
&&D_3=m_Q^2+({\bf p}_\perp + x {\bf k}_\perp)^2 \nonumber \,,
\end{eqnarray}
have been used.
Introducing the Feynman parameter $\lambda$ and performing the
integration over the transverse momentum gives
\begin{equation}
{\cal I}(\xi) =\as^2 \, \delta^{cc'} g(\xi)
\end{equation}
with
\begin{eqnarray}
\lefteqn{ g(\xi) = {1\over 2 N_c (N_c^2-1)}
\int_0^1 dx\;\int_0^1 d\lambda }\nn\\
&\times& \left[4 (2 N_c^2-1)\,\tilde{x}\,x +
    \frac{-\xi  + 2\,\tilde{x}\,x\,\left( 2 + \xi  \right) }
     {1 + (1-\lambda)\,\lambda \,\xi } +
    \frac{2 N_c^2\,x\,\left[ -4 \tilde{x}
         + x \left( {\tilde{x}}^2 + x^2 \right) \,
             \xi  \right] }{1 +
       (1-\lambda)\,x^2\,\lambda \,\xi }\right]\, .\label{intif1}
\end{eqnarray}
Note that as $\xi\to 0$, $g(\xi) \sim \ord(\xi)$, in accordance with
\eqn{QQimp3}. As $\xi\to \infty$, it grows logarithmically,
$g(\xi) \sim \log(\xi)$. The integrals of \eqn{intif1} can be performed
analytically, and we obtain
\begin{eqnarray}
\lefteqn{ g(\xi) = {1\over 9 N_c (N_c^2-1) \xi} }\label{intif2}\\
&\times& \left\{ - 4N_c^2
\left( 5\xi - 12 \right) - 3\xi + { 12 \left[ 2 N_c^2 (\xi - 2)
(\xi + 4) - \xi (\xi -1) \right]\over \sqrt{\xi(\xi +4)} }
\tanh^{-1}\sqrt{\xi\over \xi +4} \right\}\, .\nn
\end{eqnarray}

%%%%%%%%%%%%%%%%%%%%%%%%%%%%%%%%%%%%%%%%%%%%%%%%%%%%%%


\begin{thebibliography}{99}

%\cite{Frixione:1997ma}
\bibitem{Frixione:1997ma}
S.~Frixione, M.~L.~Mangano, P.~Nason and G.~Ridolfi,
{\it Heavy-quark production,}
Adv.\ Ser.\ Direct.\ High Energy Phys.\  {\bf 15} (1998) 609
[\hepph{9702287}].
%[arXiv:hep-ph/9702287].
%%CITATION = HEP-PH 9702287;%%

%\cite{Nason:1987xz}
\bibitem{Nason:1987xz}
P.~Nason, S.~Dawson and R.~K.~Ellis,
{\it The Total Cross-Section For The Production Of Heavy Quarks In Hadronic
Collisions,}
\npb{303}{1988}{607}.
%Nucl.\ Phys.\ B {\bf 303} (1988) 607.
%%CITATION = NUPHA,B303,607;%%

%\cite{Beenakker:1990ma}
\bibitem{Beenakker:1990ma}
W.~Beenakker, W.~L.~van Neerven, R.~Meng, G.~A.~Schuler and J.~Smith,
{\it QCD Corrections To Heavy Quark Production In Hadron Hadron Collisions,}
\npb{351}{1991}{507}.
%Nucl.\ Phys.\ B {\bf 351} (1991) 507.
%%CITATION = NUPHA,B351,507;%%

%\cite{Nason:1989zy}
\bibitem{Nason:1989zy}
P.~Nason, S.~Dawson and R.~K.~Ellis,
{\it The One Particle Inclusive Differential Cross-Section For Heavy Quark
Production In Hadronic Collisions,}
\npb{327}{1989}{49} [Erratum-ibid. \npb{335}{1990}{260}.
%Nucl.\ Phys.\ B {\bf 327} (1989) 49
%[Erratum-ibid.\ B {\bf 335} (1990) 260].
%%CITATION = NUPHA,B327,49;%%

%\cite{Mangano:jk}
\bibitem{Mangano:jk}
M.~L.~Mangano, P.~Nason and G.~Ridolfi,
{\it Heavy Quark Correlations In Hadron Collisions At Next-To-Leading Order,}
\npb{373}{1992}{295}.
%Nucl.\ Phys.\ B {\bf 373} (1992) 295.
%%CITATION = NUPHA,B373,295;%%

%\cite{Bonciani:1998vc}
\bibitem{Bonciani:1998vc}
R.~Bonciani, S.~Catani, M.~L.~Mangano and P.~Nason,
{\it NLL resummation of the heavy-quark hadroproduction cross-section,}
\npb{529}{1998}{424} [\hepph{9801375}].
%Nucl.\ Phys.\ B {\bf 529} (1998) 424
%[arXiv:hep-ph/9801375].
%%CITATION = HEP-PH 9801375;%%

%\cite{DelDuca:mn}
\bibitem{DelDuca:mn}
V.~Del Duca and C.~R.~Schmidt,
{\it Dijet Production At Large Rapidity Intervals,}
\prd{49}{1994}{4510} [\hepph{9311290}].
%Phys.\ Rev.\ D {\bf 49} (1994) 4510
%[arXiv:hep-ph/9311290].
%%CITATION = HEP-PH 9311290;%%

%\cite{Stirling:1994zs}
\bibitem{Stirling:1994zs}
W.~J.~Stirling,
{\it Production of jet pairs at large relative rapidity in hadron hadron
collisions as a probe of the perturbative pomeron,}
\npb{423}{1994}{56} [\hepph{9401266}].
%Nucl.\ Phys.\ B {\bf 423} (1994) 56
%[arXiv:hep-ph/9401266].
%%CITATION = HEP-PH 9401266;%%

%\cite{Andersen:2003gs}
%\bibitem{Andersen:2003gs}
%J.~R.~Andersen and W.~J.~Stirling,
%{\it Energy consumption and jet multiplicity from the leading log BFKL
%evolution,}
%\jhep{02}{2003}{018} [\hepph{0301081}].
%JHEP {\bf 0302} (2003) 018
%[arXiv:hep-ph/0301081].
%%CITATION = HEP-PH 0301081;%%

%\cite{Mueller:ey}
\bibitem{Mueller:ey}
A.~H.~Mueller and H.~Navelet,
{\it An Inclusive Minijet Cross-Section And The Bare Pomeron In QCD,}
\npb{282}{1987}{727}.
%Nucl.\ Phys.\ B {\bf 282} (1987) 727.
%%CITATION = NUPHA,B282,727;%%

%\cite{Kuraev:ge}
\bibitem{Kuraev:ge}
E.~A.~Kuraev, L.~N.~Lipatov and V.~S.~Fadin,
{\it Multi - Reggeon Processes In The Yang-Mills Theory,}
\zetf{71}{1976}{840} [\jetp{44}{1976}{443}].
%Sov.\ Phys.\ JETP {\bf 44} (1976) 443
%[Zh.\ Eksp.\ Teor.\ Fiz.\  {\bf 71} (1976) 840].
%%CITATION = SPHJA,44,443;%%

%\cite{Kuraev:1977fs}
\bibitem{Kuraev:1977fs}
E.~A.~Kuraev, L.~N.~Lipatov and V.~S.~Fadin,
{\it The Pomeranchuk Singularity In Nonabelian Gauge Theories,}
\zetf{72}{1977}{377} [\jetp{45}{1977}{199}].
%Sov.\ Phys.\ JETP{\bf 45} (1977) 199
%[Zh.\ Eksp.\ Teor.\ Fiz.\ {\bf 72} (1977) 199].
%%CITATION = ZETFA,72,377;%%

%\cite{Balitsky:1978ic}
\bibitem{Balitsky:1978ic}
I.~I.~Balitsky and L.~N.~Lipatov,
{\it The Pomeranchuk Singularity In Quantum Chromodynamics,}
\yf{28}{1978}{1597} [\sjnp{28}{1978}{822}].
%Sov.\ J.\ Nucl.\ Phys.\ {\bf 28} (1978) 822
%[Yad.\ Fiz.\ {\bf 28} (1978) 822].
%%CITATION = YAFIA,28,1597;%%

%\cite{Ellis:1990hw}
\bibitem{Ellis:1990hw}
R.~K.~Ellis and D.~A.~Ross,
{\it The Coupling Of The QCD Pomeron In Various Semihard Processes,}
\npb{345}{1990}{79}.
%Nucl.\ Phys.\ B {\bf 345} (1990) 79.
%%CITATION = NUPHA,B345,79;%%

%\cite{Stelzer:1994ta}
\bibitem{Stelzer:1994ta}
T.~Stelzer and W.~F.~Long,
{\it Automatic generation of tree level helicity amplitudes,}
Comput.\ Phys.\ Commun.\  {\bf 81} (1994) 357
[\hepph{9401258}].
%[arXiv:hep-ph/9401258].
%%CITATION = HEP-PH 9401258;%%

%\cite{Maltoni:2002qb}
\bibitem{Maltoni:2002qb}
F.~Maltoni and T.~Stelzer,
{\it MadEvent: Automatic event generation with MadGraph,}
\jhep{02}{2003}{027} [\hepph{0208156}].
%JHEP {\bf 0302} (2003) 027
%[arXiv:hep-ph/0208156].
%%CITATION = HEP-PH 0208156;%%

%\cite{Schmidt:1996fg}
\bibitem{Schmidt:1996fg}
C.~R.~Schmidt,
{\it A Monte Carlo solution to the BFKL equation,}
\prl{78}{1997}{4531} [\hepph{9612454}].
%Phys.\ Rev.\ Lett.\  {\bf 78} (1997) 4531
%[arXiv:hep-ph/9612454].
%%CITATION = HEP-PH 9612454;%%

%\cite{Orr:1997im}
\bibitem{Orr:1997im}
L.~H.~Orr and W.~J.~Stirling,
{\it Dijet production at hadron hadron colliders in the BFKL approach,}
\prd{56}{1997}{5875} [\hepph{9706529}].
%Phys.\ Rev.\ D {\bf 56} (1997) 5875
%[arXiv:hep-ph/9706529].
%%CITATION = HEP-PH 9706529;%%

%\cite{Andersen:2003an}
\bibitem{Andersen:2003an}
J.~R.~Andersen and A.~Sabio Vera,
{\it Solving the BFKL equation in the next-to-leading approximation,}
\plb{567}{2003}{116} [\hepph{0305236}].
%Phys.\ Lett.\ B {\bf 567} (2003) 116
%[arXiv:hep-ph/0305236].
%%CITATION = HEP-PH 0305236;%%

%\cite{Andersen:2003wy}
\bibitem{Andersen:2003wy}
J.~R.~Andersen and A.~Sabio Vera,
{\it The gluon Green's function in the BFKL approach at next-to-leading
logarithmic accuracy,}
\npb{679}{2004}{345} [\hepph{0309331}].
%Nucl.\ Phys.\ B {\bf 679} (2004) 345
%[arXiv:hep-ph/0309331].
%%CITATION = HEP-PH 0309331;%%

%\cite{Frixione:2002ik}
\bibitem{Frixione:2002ik}
S.~Frixione and B.~R.~Webber,
{\it Matching NLO QCD computations and parton shower simulations,}
\jhep{06}{2002}{029} [\hepph{0204244}].
%JHEP {\bf 0206} (2002) 029
%[arXiv:hep-ph/0204244].
%%CITATION = HEP-PH 0204244;%%

%\cite{Frixione:2003ei}
\bibitem{Frixione:2003ei}
S.~Frixione, P.~Nason and B.~R.~Webber,
{\it Matching NLO QCD and parton showers in heavy flavour production,}
\jhep{08}{2003}{007 } [\hepph{0305252}].
%JHEP {\bf 0308} (2003) 007
%[arXiv:hep-ph/0305252].
%%CITATION = HEP-PH 0305252;%%

%\cite{Martin:1999ww}
\bibitem{Martin:1999ww}
A.~D.~Martin, R.~G.~Roberts, W.~J.~Stirling and R.~S.~Thorne,
{\it Parton distributions and the LHC: W and Z production,}
Eur.\ Phys.\ J.\ C {\bf 14} (2000) 133 
[\hepph{9907231}].
%[arXiv:hep-ph/9907231].
%%CITATION = HEP-PH 9907231;%%

%\cite{Corcella:2000bw}
\bibitem{Corcella:2000bw}
G.~Corcella {\it et al.},
{\it HERWIG 6: An event generator for hadron emission reactions with  interfering
gluons (including supersymmetric processes),}
\jhep{01}{2001}{010} [\hepph{0011363}].
%JHEP {\bf 0101} (2001) 010
%[arXiv:hep-ph/0011363].
%%CITATION = HEP-PH 0011363;%%

%\cite{Mangano:2002wn}
\bibitem{Mangano:2002wn}
M.~L.~Mangano, M.~Moretti, F.~Piccinini, R.~Pittau and A.~D.~Polosa,
{\it b anti-b final states in Higgs production via weak boson fusion at the LHC,}
\plb{556}{2003}{50} [\hepph{0210261}].
%Phys.\ Lett.\ B {\bf 556} (2003) 50
%[arXiv:hep-ph/0210261].
%%CITATION = HEP-PH 0210261;%%

%\cite{Orr:1998hc}
\bibitem{Orr:1998hc}
L.~H.~Orr and W.~J.~Stirling,
{\it The collision energy dependence of dijet cross sections as a probe of  BFKL
physics,}
\plb{429}{1998}{135} [\hepph{9801304}].
%Phys.\ Lett.\ B {\bf 429} (1998) 135
%[arXiv:hep-ph/9801304].
%%CITATION = HEP-PH 9801304;%%

%\cite{Andersen:2001ja}
\bibitem{Andersen:2001ja}
J.~R.~Andersen, V.~Del Duca, F.~Maltoni and W.~J.~Stirling,
{\it W boson production with associated jets at large rapidities,}
%JHEP {\bf 0105} (2001) 048 [arXiv:hep-ph/0105146].
\jhep{05}{2001}{048} [\hepph{0105146}].
%%CITATION = HEP-PH 0105146;%%


\end{thebibliography}
\end{document}